\documentclass[aps,pra,
reprint,superscriptaddress,longbibliography,floatfix]{revtex4-1}
\usepackage[T1]{fontenc}
\usepackage{lmodern}
\usepackage{amsmath,amssymb,amsthm,mathtools}
\usepackage{graphicx}
\usepackage{bm}
\usepackage{booktabs}
\usepackage[hidelinks]{hyperref}
\usepackage{tikz}
\usetikzlibrary{calc}

\newcommand{\HH}{\mathcal H}
\newcommand{\SSS}{\mathfrak S}
\newcommand{\I}{\mathbb I}
\newcommand{\Tr}{\operatorname{Tr}}
\newcommand{\supp}{\operatorname{supp}}
\newtheorem{theorem}{Theorem}
\newtheorem{proposition}{Proposition}
\newtheorem{definition}{Definition}
\newtheorem{corollary}{Corollary}

\begin{document}
\title{The Structure of Shared Classical Facts in Quantum Systems}
\author{Jie Gu}
\email{jiegu1989@gmail.com}
\affiliation{Chengdu Academy of Educational Sciences, Chengdu 610036, China}
\date{\today}

\begin{abstract}
Perfectly matching local outcomes need not constitute a
nondisturbing shared classical record. For finite-dimensional
multipartite states, we prove that nontrivial exact records
are equivalent to a spectrum sharing structure with a unique
finest decomposition into locally distinguishable sectors,
where entanglement may persist.
For a fixed complete pointer, exact sharing is equivalent to
strong quantum Darwinism together with global nondisturbance,
while spectrum broadcast structure additionally requires
conditional independence.
Arbitrarily weak full-rank admixtures eliminate exact sharing,
but two-sided distance bounds quantify robustness through
disturbance and disagreement, while information bounds connect
accurate local readouts to observer consensus.
\end{abstract}
\maketitle

\section{Introduction}\label{sec:intro}

A photograph, a detector click, and a memory register may all convey the same fact even though they occupy different physical systems. Explaining how such independently accessible records emerge from quantum mechanics is a central problem in the quantum-to-classical transition. Decoherence shows how interactions with an environment suppress interference between preferred pointer alternatives in a reduced state \cite{Zurek1981,Zurek1982,JoosZeh1985,Zurek2003,Schlosshauer2005}. Yet local decoherence alone does not identify which fragments store the information, whether separate observers can access it, or whether reading it changes the larger state.

Quantum Darwinism (QD) addresses the proliferation of information into environmental fragments \cite{Ollivier2004,BlumeKohout2006,Zurek2009,Brandao2015,Zurek2022}. Its characteristic mutual-information plateau indicates that many fragments correlate strongly with the same system \cite{BlumeKohout2008,Riedel2010,Zwolak2014,Ciampini2018,Unden2019}. A plateau is not, however, a direct certificate of readable classical records: quantum mutual information also counts correlations that local measurements cannot turn into classical information about a specified pointer \cite{HendersonVedral2001,Ollivier2001,Zwolak2013,Le2018}. Strong quantum Darwinism (SQD) remedies this by requiring the relevant information to be classically accessible and excluding discord from the system--fragment correlations \cite{Le2019,Korbicz2021}.

Spectrum broadcast structure (SBS) gives a complementary state-level description \cite{Korbicz2014,Horodecki2015,Korbicz2017,Korbicz2021}. In an SBS state, distinct system-pointer values occupy orthogonal conditional supports in every observed fragment, allowing independent nondisturbing readout. SBS additionally requires those fragments to be uncorrelated conditional on the pointer value. That product assumption is stronger than the existence of a shared classical record. Indeed, SQD supplemented by a suitable multipartite independence condition yields SBS \cite{Le2019,Le2020,Feller2021}. These approaches illuminate information redundancy, local accessibility, and conditional independence, but do not make them interchangeable.

Nondisturbance is not new to structural accounts of objectivity: Bohr's requirement was used to derive the form of objective pointer records in a system--environment setting \cite{Bohr1935,Horodecki2015}. A complementary line of work uses quantum error correction and recoverable logical algebras to describe classical information available on multiple fragments \cite{Beny2007,Beny2007PRA,Girard2026}. These results motivate a distinction between the local recoverability of encoded information and a nondisturbing measurement of a particular full state. Measuring a locally represented logical observable can still dephase a coherent encoded state when its outcome is ignored.

Here we fix a multipartite density operator and a partition into disjoint observers' laboratories, without designating a privileged system or a pointer in advance. We ask whether some nonconstant classical label admits locally calibrated readouts that agree with certainty and whose outcome-forgotten action preserves the entire tested state. Both the label and the readouts are allowed to depend on that state, while the observer partition remains fixed. This state-dependent operational question differs from identifying which logical observables are representable on a code subspace, and global nondisturbance cannot be inferred from reduced system--fragment states alone.

The distinction already appears for two qubits. Both the classical mixture
\begin{equation}
\rho_{\mathrm{cl}}=\tfrac12|00\rangle\langle00|
+\tfrac12|11\rangle\langle11|
\label{eq:classicalbit}
\end{equation}
and the Bell state
$|\Phi^+\rangle=(|00\rangle+|11\rangle)/\sqrt2$
yield perfectly matching computational-basis outcomes. But only the mixture is preserved by either unread local measurement: measuring the Bell state instead removes its interbranch coherence and produces $\rho_{\mathrm{cl}}$. Agreement alone therefore does not establish an already available, nondisturbing shared fact.

Our contributions follow this operational question. First, Sec.~\ref{sec:exact} proves that exact shared records are precisely the nontrivial spectrum sharing structures: locally distinguishable orthogonal record sectors, with a unique finest decomposition of the occupied state space. Here ``spectrum'' denotes record sectors, not necessarily eigenspaces of the density operator. Conditional entanglement is allowed. Second, Sec.~\ref{sec:criteria} characterizes the extra full-state nondisturbance requirement for a prescribed complete pointer and distinguishes the resulting state family from SQD and SBS. Third, Sec.~\ref{sec:robust} derives two-sided distance bounds for a fixed local readout in terms of its disturbance and disagreement, and connects reliable readout to observer consensus. The proofs and counterexamples are collected in the appendices. Finally, Sec.~\ref{sec:discussion} discusses the physical interpretation and the open problem of record-formation dynamics.

\section{Structure of exact shared facts}\label{sec:exact}

\subsection{Two requirements for a shared record}

Consider a finite-dimensional multipartite state $\rho$ on
$\HH=\bigotimes_{a=1}^{m}\HH_a$, where $m\geq2$ and
$V=\{1,\ldots,m\}$ labels the observers. Each observer controls only
its own tensor factor. Any inaccessible environment has already been
traced out, and the partition into observers' laboratories is held
fixed throughout the test.

To ask whether one classical label is already available to everyone,
let observer $a$ perform a projective readout
$\{P_x^{(a)}\}_{x\in X}$, with $\sum_x P_x^{(a)}=\I_a$,
from a common finite set of calibrated outcomes $X$. Calibration
specifies which detector outcomes represent the same event, without
requiring identical physical detectors. Write
$\Pi_x^{(a)}=P_x^{(a)}\otimes\I_{\bar a}$ for the projector on the
complete system, where $\I_{\bar a}$ acts as the identity outside
laboratory $a$.

The readouts define an \emph{exact shared record} if (i) forgetting
the outcome of any one local measurement leaves the entire
tested state unchanged and (ii) any two observers obtain the same
calibrated answer with probability one. For projective measurements
these requirements are
\begin{equation}
\begin{aligned}
[\Pi_x^{(a)},\rho]&=0 &&\text{for every }a,x,\\
\Tr\!\left(\rho\Pi_x^{(a)}\Pi_y^{(b)}\right)&=0
&&\text{for }a\ne b,\ x\ne y.
\end{aligned}
\label{eq:operational}
\end{equation}
The first condition is equivalent to
$\sum_x\Pi_x^{(a)}\rho\Pi_x^{(a)}=\rho$ for each observer.
It concerns the full state, not only the observer's reduced state
or the probabilities of the measured outcomes. The second line gives
the probability of contradictory reports because measurements on
disjoint laboratories commute. It also implies that the probability
$p_x=\Tr(\rho\Pi_x^{(a)})$ is independent of $a$.

A \emph{nontrivial exact shared fact} is an exact shared record with
at least two outcomes of strictly positive probability. For the fixed
observer partition $V$, let $\mathrm R$ be the class of states that
admit some nontrivial exact shared record. In defining
$\mathrm R$, neither the local measurements nor their calibrated label
set is prescribed in advance.

The two requirements play different roles. The Bell state in the
Introduction gives perfectly matching $Z$ outcomes, but either unread
local measurement removes the coherence between $|00\rangle$ and
$|11\rangle$. Conversely, local measurements can preserve a state
yet produce occasional disagreements, as in the noisy example of
Sec.~\ref{sec:robust}. Exact sharing requires both nondisturbance and
agreement \cite{Bohr1935,Horodecki2015,Scandolo2021,Piani2008}.

\subsection{Locally distinguishable sectors and their characterization}

The operational conditions in Eq.~\eqref{eq:operational} have a
state-level counterpart: the state decomposes into orthogonal sectors,
and the label of each sector is locally readable in every laboratory.
For each occupied label $x$ (so $p_x>0$), define
\begin{equation}
\begin{aligned}
Q_x&=\prod_{a=1}^{m}\Pi_x^{(a)}
     =\bigotimes_{a=1}^{m}P_x^{(a)},\\
\rho_x&=\frac{Q_x\rho Q_x}{p_x},\\
\HH_{a,x}&=\operatorname{ran}P_x^{(a)}.
\end{aligned}
\label{eq:conditional-sectors}
\end{equation}
Here $Q_x$ is the joint projector for unanimous outcome $x$,
$\rho_x$ is the normalized state conditioned on that outcome, and
$\HH_{a,x}$ is the local subspace associated with $x$ in laboratory
$a$. Perfect agreement gives $p_x=\Tr(\rho Q_x)$.

The corresponding state-level condition is formalized as follows.
\begin{definition}[Spectrum sharing structure]
\label{def:spectrum-sharing}
A finite-dimensional multipartite state $\rho$ on
$\HH=\bigotimes_{a=1}^m\HH_a$ has a \emph{spectrum sharing
structure} if there exist a finite label set $X$, positive numbers
$p_x$ with $\sum_xp_x=1$, normalized states $\rho_x$, and subspaces
$\HH_{a,x}\subseteq\HH_a$ such that
\begin{equation}
\rho=\bigoplus_{x\in X}p_x\rho_x,
\qquad
\supp\rho_x\subseteq\bigotimes_{a=1}^{m}\HH_{a,x},
\label{eq:sector}
\end{equation}
and, for each observer $a$,
\begin{equation}
\HH_{a,x}\perp\HH_{a,y},\qquad x\neq y.
\label{eq:spectrum-sharing-orthogonality}
\end{equation}
The direct sum means that different labels occupy orthogonal global
blocks, with no coherence between them. The state $\rho$ has no support outside
their sum. The structure is nontrivial if $|X|\geq2$.
\end{definition}

This definition expresses the operational requirements as conditions
on the state alone. The following theorem shows that the two
descriptions are equivalent and, moreover, identifies the unique
finest decomposition into shared support sectors.

\begin{theorem}[Exact shared facts and spectrum sharing]
\label{thm:exact}
For a finite-dimensional multipartite state with at least two
disjoint observer subsystems, the following statements are equivalent:
\begin{enumerate}
\item The state admits a nontrivial exact shared record satisfying
Eq.~\eqref{eq:operational}.
\item The state has a nontrivial spectrum sharing structure in the
sense of Definition~\ref{def:spectrum-sharing}.
\end{enumerate}
Moreover, for the fixed observer partition, there is a unique
finest decomposition into shared support sectors, up to
relabeling. If $\{E_\lambda\}$ denotes the orthogonal projections
onto these nonzero sectors, then
\begin{equation}
 p_\lambda=\Tr(\rho E_\lambda),\qquad
 \rho_\lambda=\frac{E_\lambda\rho E_\lambda}{p_\lambda}.
\label{eq:finest-sectors}
\end{equation}
Every other exact shared record is obtained by combining some of
these finest labels. The projectors $E_\lambda$ act on the support
of $\rho$, so their sum is the projector onto $\supp\rho$.
When no nontrivial shared fact exists, this finest decomposition has
only one occupied sector.
\end{theorem}

Both Eqs.~\eqref{eq:sector} and
\eqref{eq:spectrum-sharing-orthogonality} are essential.
Eq.~\eqref{eq:sector} gives globally orthogonal blocks and confines
each block to the product of its local record subspaces, whereas
Eq.~\eqref{eq:spectrum-sharing-orthogonality} makes different labels
orthogonal within every individual laboratory. Thus, global
orthogonality alone does not ensure that each observer can distinguish
the sectors. Here ``spectrum'' refers to these record sectors, not
necessarily to eigenspaces of $\rho$. The conditional states $\rho_x$
need not factorize and may remain entangled across laboratories.

%For the forward implication, suppose the local readouts satisfy
%Eq.~\eqref{eq:operational}. Each unread local measurement preserves
%$\rho$, and the corresponding channels commute because they act on
%disjoint laboratories. Their product therefore leaves $\rho$
%unchanged. Perfect agreement rules out every outcome string except
%$(x,\ldots,x)$. With the joint projectors $Q_x$ defined in
%Eq.~\eqref{eq:conditional-sectors}, it follows that
%\begin{equation}
%\rho=\sum_x Q_x\rho Q_x,
%\qquad
%\sum_x\Tr(\rho Q_x)=1.
%\label{eq:operational-to-sectors}
%\end{equation}
%For every occupied label, the corresponding conditional state
%$\rho_x$ is supported in $\bigotimes_a\HH_{a,x}$. Since each local
%readout is projective, its subspaces for distinct outcomes are
%orthogonal. The decomposition therefore satisfies
%Eqs.~\eqref{eq:sector} and
%\eqref{eq:spectrum-sharing-orthogonality}.

Conversely, suppose the state has the structure in
Eqs.~\eqref{eq:sector} and
\eqref{eq:spectrum-sharing-orthogonality}. In each laboratory, measure
the orthogonal subspaces $\HH_{a,x}$, assigning any unused orthogonal
complement to one of the outcomes. The support condition ensures that
every state $\rho_x$ gives outcome $x$ with certainty in every
laboratory. The absence of coherence between different sectors then
ensures that forgetting any one outcome leaves $\rho$ unchanged.
These readouts consequently satisfy Eq.~\eqref{eq:operational}.

Finally, each exact readout induces the same projector on $\supp\rho$
in every laboratory. Projectors associated with two different readouts
can therefore be represented in different laboratories and commute.
Their common refinement gives the unique finest decomposition, and
every other exact record is a coarse-graining of it. The projector
argument is given in Appendix~\ref{app:proof}.

\subsection{Scope and implications of the characterization}
\label{sec:operational-limits}

The Bell-state example in the Introduction illustrates that
perfectly correlated local outcomes do not necessarily constitute
a nondisturbing shared fact: an unread local measurement may still
disturb the joint state. Thus, the theorem characterizes shared
records, not correlated measurement statistics alone
\cite{Ollivier2001,Luo2008,Le2018}.

A related distinction is that recovering an observable's
outcome statistics is not the same as reading an existing record
without disturbing the state. Consider the repetition encoding
$|0\rangle_L\mapsto|00\rangle$ and
$|1\rangle_L\mapsto|11\rangle$. Either physical qubit reveals
the logical $Z$ distribution. However, for an encoded superposition
\[
|\psi\rangle=\alpha|00\rangle+\beta|11\rangle,
\qquad \alpha\beta\neq0,
\]
an unread measurement of either qubit produces
\[
|\alpha|^2|00\rangle\langle00|
+
|\beta|^2|11\rangle\langle11|,
\]
destroying the coherence between the two logical branches.
Recovering the logical $Z$ statistics is therefore weaker than the exact-record criterion established here. In operator-algebra quantum error correction, local representatives of encoded observables, including central elements of recoverable algebras, are studied at the level of a code subspace \cite{Beny2007,Beny2007PRA,Girard2026}. Our test instead asks whether the actual state is invariant under an unread local projective measurement. The two notions should not be identified merely because the same classical outcome can be obtained locally. This is consistent with the no-broadcasting theorem:
the logical $Z$ label is classical, whereas an arbitrary quantum
superposition is not redundantly readable without disturbing
its coherence \cite{Barnum1996,Piani2008}.

The theorem identifies which nonconstant classical labels are
already locally readable by all observers. It does not require
the entire quantum state, or even a complete pointer basis of a
designated subsystem, to be shared. To compare this general
shared-record criterion with frameworks based on a designated
system pointer, Sec.~\ref{sec:criteria} fixes such a pointer
and examines its relation to quantum Darwinism and spectrum
broadcast structure.

\section{Shared facts and quantum objectivity}\label{sec:criteria}\label{sec:pointerhierarchy}

For comparison with quantum Darwinism, fix one finite-dimensional
state $\rho_{SF_1\cdots F_m}$, a designated system $S$, and $m\geq2$
disjoint environmental fragments $F_a$. Abbreviate the collection of
fragments as $\bm F=F_1\cdots F_m$. For our shared-record test, $S$
and each $F_a$ count as separate observers, so
$V=\{S,F_1,\ldots,F_m\}$. All exact criteria below refer to the
same full state and this fixed partition. Write $H(A)$ for
von Neumann entropy and $I(A:B)=H(A)+H(B)-H(AB)$ for quantum mutual
information. For a transparent fixed-partition comparison, we use the idealized exact equality below as the QD criterion for every displayed fragment. This is not the same as testing an approximate or fragment-size-averaged mutual-information plateau in a dynamical model. Conventional quantum Darwinism (QD) is represented here by
\begin{equation}
\mathrm{QD}:\quad I(S:F_a)=H(S)\qquad(\forall a).
\label{eq:qdtest}
\end{equation}
To compare complete-pointer records, fix a rank-one orthonormal
basis $X=\{|x\rangle_S\}_x$ of $S$ and restrict the state families
to states whose $\rho_S$ is diagonal in this basis. Thus $X$ is an
eigenbasis of $\rho_S$ (including any degenerate eigenspaces),
$p_x=\langle x|\rho_S|x\rangle$, and only labels with $p_x>0$
need be distinguished. We consider states with at least two occupied
pointer labels ($H(S)>0$), consistently with the nontrivial-record
convention of Sec.~\ref{sec:exact}. These assumptions are important
when translating entropic criteria into an exact state form and
comparing the state families below.
Strong quantum Darwinism (SQD) requires all the correlation to be
accessible as classical information about this complete pointer:
\begin{equation}
\begin{split}
\mathrm{SQD}_X:\quad I(S:F_a)&=I_{\mathrm{acc}}(S:F_a)\\
&=\chi(X:F_a)=H(S)\quad(\forall a).
\end{split}
\label{eq:sqdtest}
\end{equation}
Here $I_{\mathrm{acc}}(S:F_a)$ maximizes the classical mutual
information over local measurements on $S$ and $F_a$. The Holevo
quantity is $\chi(X:F_a)=H(\rho_{F_a})-
\sum_{x:p_x>0}p_xH(\tau_x^{(a)})$, with
$\tau_x^{(a)}=p_x^{-1}\langle x|\rho_{SF_a}|x\rangle_S$
the conditional fragment state \cite{Holevo1973}. We use
$H(X)=-\sum_{x:p_x>0}p_x\log_2p_x$ for the Shannon entropy of
the pointer outcomes. By the diagonal-pointer assumption,
$H(X)=H(S)$. Under this assumption, ideal SQD is equivalent to the
pairwise state form
\begin{equation}
\begin{aligned}
\rho_{SF_a}&=\sum_xp_x|x\rangle\langle x|_S\otimes\tau_x^{(a)},\\
\tau_x^{(a)}\tau_y^{(a)}&=0\quad(x\neq y,\ p_xp_y>0).
\end{aligned}
\label{eq:pairwisebroadcast}
\end{equation}
Thus each pairwise $S:F_a$ state records the full pointer, but none of these marginals by itself tests whether the complete $S\bm F$ state retains coherence between distinct pointer branches \cite{Le2019,Korbicz2021}. Classical accessibility in each pairwise marginal and nondisturbance of the global state are therefore separate requirements. The equivalence between the entropic and state-based forms is derived in Appendix~\ref{app:pointerproof}.

Spectrum broadcast structure (SBS) further requires independence of
the fragments conditioned on the pointer value:
\begin{equation}
\begin{split}
\rho_{S\bm F}^{\rm SBS}
&=\sum_xp_x|x\rangle\langle x|_S
       \otimes\bigotimes_{a=1}^{m}\tau_x^{(a)},\\
\tau_x^{(a)}\tau_y^{(a)}&=0\quad(x\neq y,\ p_xp_y>0).
\end{split}
\label{eq:sbscompare}
\end{equation}
The standard SQD--SBS relation is therefore an implication from
conditional product structure to locally accessible information
\cite{Le2019,Le2020,Feller2021}.

For this fixed partition and complete pointer $X$, let
$\mathrm R_X$ denote the class of states for which the prescribed
rank-one projective measurement
$\{|x\rangle\langle x|_S\}_x$ and matching, calibrated
projective readouts on every fragment $F_a$ satisfy both conditions
of Eq.~\eqref{eq:operational} on the full state. Thus
$\mathrm R_X$ requires nondisturbing access to every
occupied value of the specified complete system pointer.
By contrast, $\mathrm R$ requires only some nonconstant
exact shared event, which may be a coarser observable of $S$ and
need not correspond to the chosen pointer $X$. The definition of
$\mathrm R$ does not single out a designated system. Equivalently, it only
requires a nontrivial spectrum sharing structure by
Theorem~\ref{thm:exact}. Consequently,
$\mathrm R_X\subseteq\mathrm R$ on this fixed observer partition,
but a state in $\mathrm R$ need not belong to $\mathrm R_X$.
All comparisons between $\mathrm{SQD}_X$, $\mathrm R_X$,
and $\mathrm{SBS}_X$ use this fixed pointer and the
diagonal-pointer state family.

\begin{proposition}[Exact pointer-record overlap]\label{prop:overlap}
For this nontrivial complete rank-one pointer $X$ and fixed fragments,
\begin{equation}
\mathrm R_X=\mathrm{SQD}_X\cap\mathrm{ND}_X
\label{eq:overlapidentity}
\end{equation}
where $\mathrm{ND}_X$ means
$[|x\rangle\langle x|_S\otimes\I_{\bm F},\rho]=0$
for each occupied $x$. Equivalently, the global state has the form
\begin{equation}
\begin{aligned}
\rho_{S\bm F}&=\sum_xp_x|x\rangle\langle x|_S\otimes\tau_x,\\
\supp\tau_x&\subseteq\bigotimes_a\HH_{F_a,x},\\
\HH_{F_a,x}&\perp\HH_{F_a,y}\quad(x\ne y,\ p_xp_y>0).
\end{aligned}
\label{eq:generalpointer}
\end{equation}
No factorization of $\tau_x$ is required. Within this family SBS
holds precisely when
$\tau_x=\bigotimes_a\tau_x^{(a)}$ for each occupied label.
\end{proposition}

The proposition (proved in Appendix~\ref{app:pointerproof}) separates local readability from global nondisturbance: $\mathrm{SQD}_X$ certifies the former in the $S:F_a$ marginals, while $\mathrm{ND}_X$ excludes coherence between pointer sectors in the full tested state. Earlier identifications of SQD with pointer objectivity \cite{Le2019} are thus not being contradicted: the present criterion explicitly tests a stronger, full-state outcome-forgotten readout on a specified multipartite partition. For instance, a coherent GHZ state has the ideal SQD pairwise marginals, although reading any one pointer record dephases the complete pure state (Example C in Appendix~\ref{app:counterexamples}). The fixed-pointer hierarchy is therefore
\begin{equation}
\mathrm{SBS}_X\subsetneq\mathrm R_X
\subsetneq\mathrm{SQD}_X\subseteq\mathrm{QD}.
\label{eq:pointerhierarchy}
\end{equation}
The general shared-event class $\mathrm R$ is broader than $\mathrm R_X$ and has no general inclusion relation with QD or SQD.

The strict distinction between $\mathrm R_X$ and SBS is already visible in a simple example. Let $F_a=L_aK_a$, where each $L_a$ holds an orthogonal copy of a pointer bit, and let the two $K$ registers share $|\phi_\theta\rangle=\cos\theta|00\rangle+\sin\theta|11\rangle$. The state
\[
\begin{aligned}
\rho_D(\theta)={}&\frac12\sum_{x=0}^{1}|x\rangle\langle x|_S
\otimes|xx\rangle\langle xx|_{L_1L_2}\\
&\otimes|\phi_\theta\rangle\langle\phi_\theta|_{K_1K_2}
\end{aligned}
\]
has perfectly readable, nondisturbing pointer copies in $L_1$ and $L_2$. Yet for $0<\theta\leq\pi/4$, the conditional $K_1K_2$ state is entangled, so the fragments are not conditionally independent and the state is not SBS. The explicit verification is in Example D of Appendix~\ref{app:counterexamples}. Figure~\ref{fig:framework} summarizes the full hierarchy and its other witnesses.

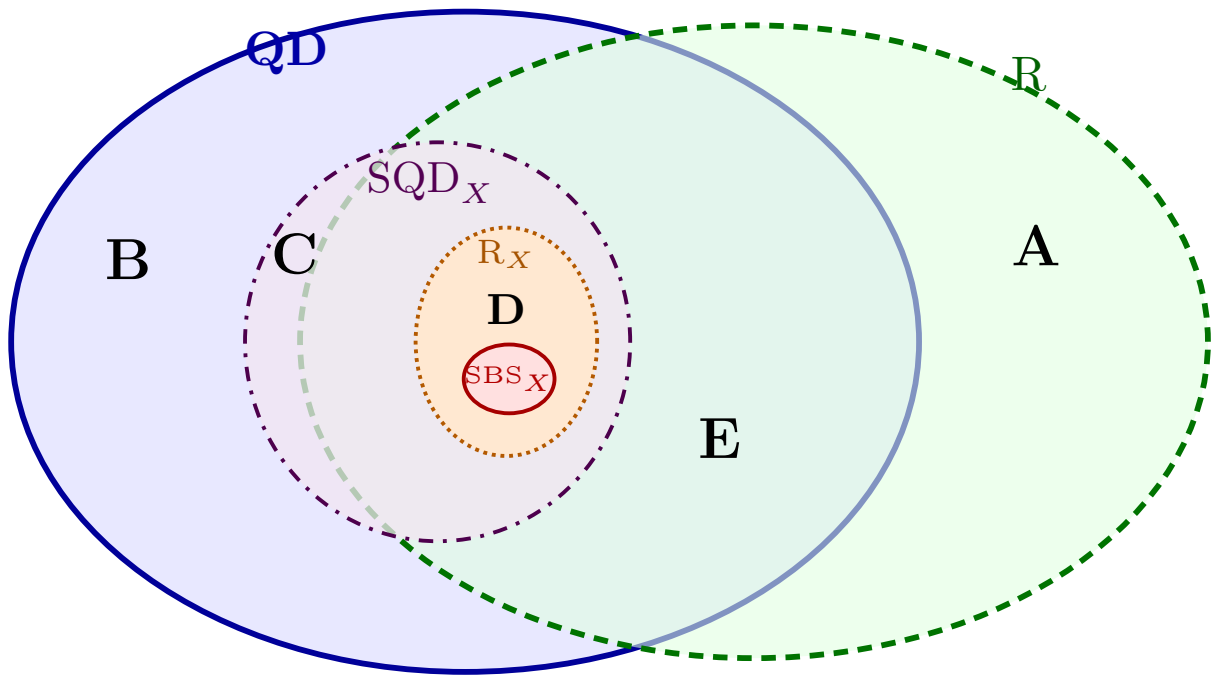
\begin{figure}[t]
\centering
\resizebox{0.92\columnwidth}{!}{%
\begin{tikzpicture}[font=\small, line join=round]
  % The geometry encodes SBS_X \subset R_X \subset SQD_X \subset QD,
  % with R intersecting QD and extending outside it.
  \filldraw[fill=blue!9,draw=blue!60!black,very thick]
      (4,3) ellipse [x radius=3.3,y radius=2.4];
  \filldraw[fill=green!12,fill opacity=0.58,
      draw=green!45!black,very thick,densely dashed]
      (6.1,3) ellipse [x radius=3.3,y radius=2.3];
  \filldraw[fill=violet!10,fill opacity=0.75,
      draw=violet!60!black,thick,dash dot]
      (3.8,3) ellipse [x radius=1.4,y radius=1.45];
  \filldraw[fill=orange!18,draw=orange!70!black,thick,densely dotted]
      (4.3,3) ellipse [x radius=0.66,y radius=0.83];
  \filldraw[fill=red!12,draw=red!65!black,thick]
      (4.32,2.73) ellipse [x radius=0.33,y radius=0.25];
  \node[font=\bfseries\normalsize,text=blue!65!black]
      at (2.7,5.1) {QD};
  \node[font=\bfseries\normalsize,text=green!38!black]
      at (8.10,4.95) {$\mathrm R$};
  \node[font=\bfseries\small,text=violet!65!black]
      at (3.75,4.16) {$\mathrm{SQD}_X$};
  \node[font=\bfseries\scriptsize,text=orange!65!black]
      at (4.3,3.64) {$\mathrm R_X$};
  \node[font=\bfseries\tiny,text=red!70!black]
      at (4.32,2.73) {$\mathrm{SBS}_X$};
  \node[font=\bfseries\large] at (8.15,3.70) {A};
  \node[font=\bfseries\large] at (1.55,3.60) {B};
  \node[font=\bfseries\large] at (2.76,3.64) {C};
  \node[font=\bfseries\small] at (4.30,3.24) {D};
  \node[font=\bfseries\large] at (5.85,2.30) {E};
\end{tikzpicture}%
}
\caption{\textbf{Fixed-partition hierarchy and its generalization.}
The outer Euler sets compare exact shared facts $\mathrm R$ with QD
and $\mathrm{SQD}_X$. Distinct boundary styles also distinguish the
sets in grayscale reproduction. The nested sets distinguish
$\mathrm{SBS}_X\subsetneq\mathrm R_X\subsetneq\mathrm{SQD}_X$.
Here A denotes an incomplete shared label, B two Bell pairs,
C a coherent GHZ state, D an exact pointer record with conditional
entanglement, and E a shared tag accompanying two Bell pairs.
The state families are compared for a common observer partition
and, when specified, the same complete pointer $X$. The diagram does
not represent a random-partition consensus probability.}
\label{fig:framework}
\end{figure}

Examples A--E in Fig.~\ref{fig:framework} show how the boundaries
differ. A and B separate general shared records from QD in both
directions. C satisfies SQD but fails global nondisturbance. D is an
exact pointer record that is not SBS. E belongs to
$\mathrm R\cap\mathrm{QD}$ but not SQD.

Conditional correlations explain the gap between $\mathrm R_X$ and
SBS. For the conditional states $\tau_x$ in
Eq.~\eqref{eq:generalpointer}, define
\begin{equation}
\begin{split}
J(F_1:\cdots:F_m|S)
 &=\sum_xp_x\!\left[\sum_a H(\tau_x^{(a)})-H(\tau_x)\right]\\
 &=\sum_xp_xD\!\left(\tau_x\middle\|
       \bigotimes_a\tau_x^{(a)}\right)\geq0,
\end{split}
\label{eq:conditionaltotal}
\end{equation}
where $D(\alpha\|\beta)=\Tr[\alpha(\log_2\alpha-\log_2\beta)]$
is quantum relative entropy, with the usual support convention.
Within $\mathrm R_X$, $J=0$ if and only if the state is SBS.
Otherwise, conditional correlations, including entanglement, may
remain despite exact local access to the pointer.

These comparisons depend on the tested state and the chosen observer
partition. For fixed disjoint laboratories, a record shared by all
members of a group is also shared by any subgroup with at least two
observers. Adding observers can only restrict the set of shared
records. We assume local projective readouts and nondisturbance of
the full state, rather than arbitrary decoding instruments.
Randomly sampled fragment partitions are considered separately in
Sec.~\ref{sec:consensus}.

\section{From exact shared facts to robust records}\label{sec:robust}

\subsection{Exactness and rank-changing noise}

Suppose the shared classical bit in Eq.~\eqref{eq:classicalbit} is contaminated by a small amount of completely random two-qubit noise:
\begin{equation}
\rho_\eta=(1-\eta)\rho_{\mathrm{cl}}+\eta\I_4/4,
\qquad 0\leq\eta\leq1.
\label{eq:noise}
\end{equation}
Here $\eta$ is the noise strength. With probability $1-\eta$ we prepare the perfectly matched bit, while with probability $\eta$ we prepare a uniformly random two-bit string. The probability of disagreement is therefore $\eta/2$. Both observers can still read in the computational basis without disturbing the state, because it is diagonal in that basis.

Yet the exact-record property changes abruptly. At $\eta=0$, the two observers share one exact bit. For every $\eta>0$, even if extremely small, $\rho_\eta$ has nonzero weight on all four computational-basis strings. In particular, mismatched outcomes are now possible, so no nontrivial exact shared fact remains. A more general statement follows.

\begin{corollary}[No nontrivial exact shared fact at full rank]\label{cor:fullrank}
If a multipartite state has full rank on the tensor product of two or more disjoint observer subsystems, then it admits no nontrivial exact shared record for that observer partition.
\end{corollary}

The reason is elementary. At full rank, every pair of nonzero local projectors on distinct laboratories has strictly positive joint probability. If two different shared labels both occur, some pair of mismatched outcomes must then have positive probability, contradicting Eq.~\eqref{eq:operational}. A proof is given in Appendix~\ref{app:proof}. When the support is smaller, several different local projectors may agree on the part of Hilbert space that is actually occupied.

It would be misleading to conclude that a tiny $\eta$ suddenly erases a macroscopic memory. A detector with a one-in-a-million disagreement probability is nearly as useful as a perfect detector, despite failing an exact all-or-nothing test. To describe real records, we therefore keep the proposed local measurements fixed and quantify their imperfections.

\subsection{Operational errors and distance to ideal records}

Let each observer attempt to read the same finite label $x$ using projectors $\{P_x^{(a)}\}_{x\in X}$. We do not assume that these measurements are nondisturbing or that they agree. Two directly interpretable errors are enough for our bound.

First, imagine that all observers make their measurements and then throw away the recorded outcomes. The resulting state is
\begin{equation}
\Delta(\rho)=\sum_{\bm x}\Pi_{\bm x}\rho\Pi_{\bm x},\qquad
\Pi_{\bm x}=\bigotimes_{a=1}^m P_{x_a}^{(a)}.
\label{eq:pinching}
\end{equation}
The tuple $\bm x=(x_1,\ldots,x_m)$ lists the outcomes in all laboratories. The map $\Delta$ removes coherence between alternatives distinguished by the complete set of local measurements. We quantify its disturbance by $\delta=\|\rho-\Delta(\rho)\|_1$.

Second, let $Q_x=\bigotimes_aP_x^{(a)}$ project onto the event that every observer reports the same particular value $x$, and let $Q=\sum_xQ_x$ collect all such events. The disagreement probability is $\epsilon=1-\Tr(Q\rho)$. In summary,
\begin{equation}
\delta=\|\rho-\Delta(\rho)\|_1,\qquad
\epsilon=1-\Tr(Q\rho).
\label{eq:errors}
\end{equation}

For a Hermitian operator, $\|\cdot\|_1$ means the sum of the absolute values of its eigenvalues. We use the unhalved trace norm throughout. The conventional trace distance between density operators is often defined with an extra factor $1/2$. Thus $\delta=0$ means no disturbance, and $\epsilon=0$ means perfect agreement. The noisy shared bit has $\delta=0$ and $\epsilon=\eta/2$.

What do small values of these errors guarantee? Define $\SSS_Q$ as the family of ideal states that give only matching local outcomes and have no coherence between different values of the shared label:
\begin{equation}
\SSS_Q=\left\{\sigma\geq0:\ \Tr\sigma=1,\quad
\sigma=\sum_x Q_x\sigma Q_x\right\}.
\label{eq:idealset}
\end{equation}
For this specified choice of local measurements, the quantity
\begin{equation}
d_Q(\rho)=\inf_{\sigma\in\SSS_Q}\|\rho-\sigma\|_1
\label{eq:distance}
\end{equation}
is the distance to the nearest ideal shared-record state. It asks how much the physical state would need to change before the proposed records became perfect. The optimization does not alter which measurements the observers are allowed to perform.

\begin{theorem}[Accuracy of a fixed shared readout]\label{thm:robust}
For any family of local projective measurements with $\epsilon<1$, the distance to the ideal shared-record states satisfies
\begin{equation}
\max\{\delta/2,\,2\epsilon\}\ \leq\
d_Q(\rho)\ \leq\ \delta+2\epsilon.
\label{eq:twosided}
\end{equation}
In particular, both errors vanish exactly if and only if $\rho\in\SSS_Q$.
\end{theorem}

This bound has a useful reading without the proof. If the unread measurement strongly changes the state, or if the observers frequently disagree, the state cannot be close to a perfect shared record. This is the lower bound. Conversely, when both effects are small, there is an ideal state nearby that makes the same local projectors into a perfect, nondisturbing record. This is the upper bound.

The upper estimate can be understood as a two-step repair procedure. First perform the unread local measurement, producing $\Delta(\rho)$. This costs at most $\delta$ in trace norm. Then keep only the subspace in which all observers agree and normalize the remaining state. Because the first step has already removed coherence between different outcome strings, this second correction costs exactly $2\epsilon$, not a term proportional to $\sqrt\epsilon$. Appendix~\ref{app:trace} gives the proof and explains why the coefficient cannot be improved in general.

This result has a different target from earlier geometric bounds on proximity to spectrum broadcast structure, which use pointer dephasing and distinguishability of conditional environmental states \cite{Mironowicz2017,Le2019}. The ideal set $\SSS_Q$ here fixes the observers' proposed projectors but does not require conditional independence of the fragments. Moreover, Eq.~\eqref{eq:twosided} supplies both a lower and an upper estimate using directly interpretable errors of that same readout. It therefore quantifies nondisturbing agreement rather than proximity to the more restrictive SBS family.

For the noisy bit in Eq.~\eqref{eq:noise}, the bound is particularly transparent:
\begin{equation}
\delta=0,\qquad\epsilon=\eta/2,\qquad d_Q(\rho_\eta)=\eta.
\label{eq:noiseresult}
\end{equation}
The distance from a perfect shared record goes smoothly to zero as the noise vanishes, even though no exact shared fact remains for any positive $\eta$.

The errors are also stable if the prepared state changes slightly while the detector settings remain fixed. If two states are separated by an unhalved trace-norm distance $s$, their disturbance parameters differ by at most $2s$ and their disagreement probabilities by at most $s/2$. Thus the easily evaluated expression $\delta+2\epsilon$ changes by no more than $3s$. The estimates follow directly from the fact that quantum measurements cannot increase trace distance. Details are in Appendix~\ref{app:trace}. We are not claiming that the identity of an optimal detector remains unchanged when it is reoptimized after each perturbation.

\subsection{Accessible information and observer consensus}\label{sec:consensus}

The fixed-pointer inclusion in Eq.~\eqref{eq:pointerhierarchy} concerns
exact states for one partition. The errors above also give
one-sided statements for approximate records. Suppose the specified
local measurements include a complete rank-one pointer $X$ on $S$.
Let $Y_a$ be the calibrated outcome at fragment $F_a$ and
$e_a=\Pr(Y_a\neq X)\leq\epsilon$. Write
$h_2(u)=-u\log_2u-(1-u)\log_2(1-u)$, with $0\log_20=0$,
for binary entropy. For $k$ pointer labels, Fano's inequality \cite{Fano1961} yields
\begin{equation}
\begin{aligned}
I_{\rm acc}(S:F_a)&\geq I(X:Y_a)\\
&\geq H(X)-h_2(e_a)-e_a\log_2(k-1).
\end{aligned}
\label{eq:fanobridge}
\end{equation}
If $\rho_S$ is diagonal in $X$, then $H(X)=H(S)$.
For $\epsilon\leq1-1/k$, $e_a$ may be replaced by $\epsilon$ in the
lower bound. The qualification about the full pointer matters:
an incomplete tag as in Eq.~\eqref{eq:exampleA} of Appendix~\ref{app:counterexamples} is not enough to
certify SQD of the entire system.

There is also a geometric implication. Let $\mathsf{SQD}_X$ denote
the global states for which every $S:F_a$ marginal has the
ideal form of Eq.~\eqref{eq:pairwisebroadcast}. Every ideal
complete-pointer record belongs to this family, so Theorem~\ref{thm:robust}
gives
\begin{equation}
\inf_{\omega\in\mathsf{SQD}_X}\|\rho-\omega\|_1
\ \leq\ d_Q(\rho)\ \leq\ \delta+2\epsilon.
\label{eq:sqdgeometric}
\end{equation}
The converse fails for the coherent GHZ state $\rho_C$ of
Appendix~\ref{app:counterexamples}: its SQD marginals are ideal,
but the unread $Z$ measurement gives $\delta=1$, $\epsilon=0$,
and $d_Q(\rho_C)=1$. Conversely, $d_Q(\rho_D(\theta))=0$ in
that appendix even when the fragments are conditionally entangled,
so exact sharing does not require SBS.

The question then shifts from which event is shared for a
fixed cover to how often different covers reveal it.
Following the distinction between redundancy (existence of a suitable
partition) and consensus (success under a distribution of partitions)
\cite{Chisholm2023,Chisholm2024}, let $\mathfrak P_m$ be a family of
partitions of the environment into $m$ disjoint equally sized
fragments, sampled from a probability distribution $\mu_m$. For an
allowed accessible-information deficit $\zeta\geq0$, define the
success probability
\begin{equation}
P_{\rm acc}(\rho,m,\zeta)=\Pr_{\pi\sim\mu_m}\!
\left[\forall a,\ I_{\rm acc}(S:F_a^\pi)\geq H(S)-\zeta\right].
\label{eq:Pacc}
\end{equation}
For the same complete pointer and tested full state, fix a
nonnegative error tolerance $t$ and require local readouts to satisfy
$\delta_\pi+2\epsilon_\pi\leq t$. Define
\begin{equation}
\begin{split}
P_{\rm rec}(\rho,m,t)=\Pr_{\pi\sim\mu_m}\!\big[\,
&\exists\text{ local readouts}:\\
&\delta_\pi+2\epsilon_\pi\leq t\,\big].
\end{split}
\label{eq:Prec}
\end{equation}
Readouts may be optimized separately for each partition, but the
pointer, full state, and sampling distribution remain fixed.

\begin{proposition}[Shared-record success bounds accessible-information consensus]\label{prop:consensus}
Let $X$ be a complete rank-one eigenbasis of $\rho_S$ with $k\geq2$
labels, and let $0\leq t\leq2(1-1/k)$. With
\begin{equation}
f_k(u)=h_2(u)+u\log_2(k-1),
\label{eq:fanof}
\end{equation}
any distribution over the partitions satisfies
\begin{equation}
P_{\rm rec}(\rho,m,t)\leq
P_{\rm acc}\bigl(\rho,m,f_k(t/2)\bigr).
\label{eq:probabilitybridge}
\end{equation}
The same one-way implication holds for the largest successful
partition size and for any prescribed probability threshold.
\end{proposition}

The proof in Appendix~\ref{app:pointerproof} applies the Fano bound
to each successful partition before averaging over $\mu_m$.
It uses no conditional independence. The reverse implication is false:
the GHZ state $\rho_C$ of Appendix~\ref{app:counterexamples}
has full accessible pointer information on its natural partition,
yet an unread measurement disturbs its global coherence. This is why a fixed-partition hierarchy must not be
confused with a random-partition consensus score. A partition engineered
to expose the available records may be rare under a different sampling
distribution \cite{Chisholm2023,Chisholm2024}.

Finally, the number of possible shared labels is not the same as the information usually carried by them. If the finest shared-sector decomposition contains $n$ occupied labels, a uniformly distributed sector label carries $\log_2n$ bits of Shannon information. The actual average information is the Shannon entropy $H(X)=-\sum_xp_x\log_2p_x$. A record with many possible outcomes may carry little information if one outcome occurs almost every time. For imperfect readouts it is therefore useful to report the outcome probabilities together with $\delta$ and $\epsilon$ \cite{Groisman2005,Fuchs1999}.

\section{Discussion and outlook}\label{sec:discussion}\label{sec:branch}

The state-level results obtained here have three complementary parts. Theorem~\ref{thm:exact} gives a unique finest locally readable record-sector structure without prescribing a pointer. Proposition~\ref{prop:overlap} locates complete-pointer sharing relative to SQD and SBS by adding a full-state nondisturbance condition. Theorem~\ref{thm:robust} bounds approximate sharing through disturbance and disagreement. These conclusions neither require conditional fragment independence nor identify accessible information with nondisturbing readout. In particular, the loss of exact sharing under full-rank noise need not imply an abrupt loss of practical reliability, and Proposition~\ref{prop:consensus} relates reliable records to consensus only in one direction.

These statements concern the properties of a state at a given time,
not the interaction that prepared it. In the usual decoherence and
quantum Darwinism picture, a system pointer becomes correlated with
the environment \cite{Zurek2003,Ollivier2004,Zurek2009,Zwolak2016}, schematically as
\begin{equation}
|\Psi\rangle=\sum_x c_x|x\rangle_S|E_x\rangle.
\label{eq:branchstate}
\end{equation}
Tracing out inaccessible degrees of freedom may suppress interference
between pointer values, but this does not by itself ensure that the
fragments available to separate observers can distinguish those
values. Suppression of pointer coherence and local readability are
distinct requirements \cite{GarciaPerez2020,Lee2024Complement,Roszak2019}.

A useful bridge from this physical picture to approximate records can
be stated without specifying a dynamical model. Let $\rho_{S\bm F}$
be the state of the system and its accessible fragments, and define
the system-pointer dephasing map by
\[
\Delta_S(\rho)=\sum_x\Pi_x^{(S)}\rho\Pi_x^{(S)},\qquad
\Pi_x^{(S)}=|x\rangle\langle x|_S\otimes\I_{\bm F}.
\]
The residual pointer coherence is
\begin{equation}
D_{\mathrm{coh}}=\|\rho_{S\bm F}-\Delta_S(\rho_{S\bm F})\|_1.
\label{eq:pointerD}
\end{equation}
For orthogonal fragment readout projectors $\{P_x^{(a)}\}_x$, set
$Q_x=|x\rangle\langle x|_S\otimes\bigotimes_aP_x^{(a)}$ and
$Q=\sum_xQ_x$. The probability that at least one fragment fails
to identify the pointer after dephasing is
$r=1-\Tr[Q\Delta_S(\rho_{S\bm F})]$. If $r<1$, then
$\sigma=Q\Delta_S(\rho_{S\bm F})Q/(1-r)$ is an ideal shared-record
state for these readouts, and
\begin{equation}
\|\rho_{S\bm F}-\sigma\|_1\leq D_{\mathrm{coh}}+2\sqrt r+r.
\label{eq:branch}
\end{equation}
Thus small residual pointer coherence and small fragment-identification
error suffice to place the state close to an ideal shared record,
without requiring conditional independence. The square-root term
accounts for the possible disturbance caused by projecting the
fragments onto the matching-outcome subspace. A proof is given in
Appendix~\ref{app:trace}.

If $p_x$ denotes the pointer probabilities and $\rho_x^{(a)}$ the
conditional state of fragment $a$, the mean local identification error is
\[
r_a=\sum_xp_x\Tr\!\left[(\I-P_x^{(a)})\rho_x^{(a)}\right].
\]
The union bound gives
\begin{equation}
r\leq\sum_a r_a,
\label{eq:unionbound}
\end{equation}
regardless of correlations or entanglement between fragments.
This estimate and Eq.~\eqref{eq:branch} complement, rather than replace, the fixed-readout bound in Eq.~\eqref{eq:twosided}: the former uses pointer coherence and fragment-identification errors, whereas the latter uses the disturbance $\delta$ of the complete local readout and its disagreement probability $\epsilon$. These are state-level sufficient conditions, not equations governing the time evolution of records.

Neither bound determines how quickly records form. In particular,
decoherence of the system pointer alone need not make information
accessible to individual fragments \cite{GarciaPerez2020,Lee2024Complement,Lorenzo2020,Girolami2022,Gu2026}.
In environments with memory, the quality of a record can vary with
time \cite{Riedel2012,Pleasance2017,Lampo2017,Mironowicz2018}. For fixed detectors,
$\delta(t)$ and $\epsilon(t)$ can track such changes, but
nondisturbance at one instant does not establish nondisturbance in
a multitime experiment \cite{Sakuldee2022,Strasberg2023}.

A natural direction for future work is therefore a dynamical theory
of shared-record formation under specified system--environment
interactions \cite{Tuziemski2015,Tuziemski2016,Mironowicz2017,Campbell2019,Doucet2024,Ferte2024}. Such a study could track $D_{\mathrm{coh}}(t)$,
$r(t)$, $\delta(t)$, and $\epsilon(t)$ for physically motivated
fragment partitions, determine when local readability develops
relative to decoherence, and investigate record-formation rates,
persistence, and memory-dependent effects. These are questions about
the production and temporal stability of records, distinct from the
state-level existence and robustness criteria established here.
The present results provide a quantitative benchmark for addressing
them.

\begin{acknowledgments}
AI-assisted scientific reasoning, literature organization and drafting were used in preparing this manuscript with OpenAI GPT-6.
The author is responsible for all the contents.
\end{acknowledgments}

\appendix

\section{Proof of the shared-sector structure}\label{app:proof}

\subsection{Proof of the spectrum sharing characterization}

Let the local projective measurements satisfy
Eq.~\eqref{eq:operational}. Since $[\Pi_x^{(a)},\rho]=0$, sequential
unread measurements in the different laboratories leave $\rho$
unchanged. Their projectors commute because the laboratories are
disjoint, so the resulting state is
\begin{equation}
\rho=\sum_{x_1,\ldots,x_m}
\left(\bigotimes_a P_{x_a}^{(a)}\right)\rho
\left(\bigotimes_a P_{x_a}^{(a)}\right).
\end{equation}
If the outcome string has two unequal entries, its projector is
bounded by the product of the corresponding two local projectors.
The probability of that string therefore vanishes by the second
condition of Eq.~\eqref{eq:operational}. A positive density matrix
with zero probability on a projector has no support on that
projector. Only the strings $(x,\ldots,x)$ remain, which gives
the block decomposition in Eq.~\eqref{eq:sector}. For each occupied label $x$,
put
\begin{equation}
 p_x=\Tr(\rho Q_x),\qquad
 \rho_x=\frac{Q_x\rho Q_x}{p_x},\qquad
 \HH_{a,x}=\operatorname{ran}P_x^{(a)}.
\end{equation}
The $\HH_{a,x}$ are orthogonal as $x$ varies, and
$\supp\rho_x\subseteq\bigotimes_a\HH_{a,x}$.
This establishes the nontrivial spectrum sharing structure whenever
the original record is nontrivial.

Conversely, suppose $\rho$ has a nontrivial decomposition satisfying
both Eqs.~\eqref{eq:sector} and
\eqref{eq:spectrum-sharing-orthogonality}. Let $R_{a,x}$ be the
orthogonal projector onto $\HH_{a,x}$. Because these subspaces are
pairwise orthogonal,
$R_{a,\perp}=\I_a-\sum_xR_{a,x}$ is also a
projector. Fix one label $x_0$ and define a complete local readout by
\begin{equation}
P_x^{(a)}=
\begin{cases}
R_{a,x}+R_{a,\perp},&x=x_0,\\
R_{a,x},&x\ne x_0.
\end{cases}
\label{eq:proof-sector-readout}
\end{equation}
The arbitrary assignment of unused local subspaces cannot affect
any outcome on the support of $\rho$. For all $x,y$,
\begin{equation}
\Pi_x^{(a)}\rho_y=\rho_y\Pi_x^{(a)}
=\delta_{xy}\rho_y.
\label{eq:proof-sector-action}
\end{equation}
It follows that every local measurement commutes with the full
$\rho$ and that all observers obtain the same label with certainty.
Because at least two $p_x$ are positive, the resulting record is
nontrivial. This completes the proof of the principal equivalence
without using operator algebras.

\subsection{Uniqueness of the finest shared-sector decomposition}

We now prove the uniqueness claim without introducing an operator
algebra. Let $P$ project onto $\supp\rho$. Any local projector
$\Pi_x^{(a)}$ of an exact shared readout commutes with $\rho$ and
therefore with $P$. By the first part of the proof, the occupied
part of the state lies entirely in the matching-outcome subspaces
$Q_x=\prod_a\Pi_x^{(a)}$. Hence
\begin{equation}
\Pi_x^{(a)}P=Q_xP\equiv E_x
\qquad\text{for every observer }a.
\label{eq:common-compression}
\end{equation}
Thus every observer implements the same projector $E_x$ on the
occupied state space, even if their physical projectors differ on
subspaces that are never occupied. The $E_x$ are orthogonal
projections summing to $P$.

Consider two possibly different exact shared readouts, with
projectors $E_x$ and $F_y$ on the support. Choose different observers
$a$ and $b$. According to Eq.~\eqref{eq:common-compression}, $E_x$
can be represented on the support by a local projector in laboratory
$a$, and $F_y$ by one in laboratory $b$. Local projectors acting on
disjoint tensor factors commute. Since both commute with $P$, we get
\begin{equation}
E_xF_y=F_yE_x.
\label{eq:compatible-refinements}
\end{equation}
This argument applies to every pair of projectors belonging to
any exact shared readouts, not just to projectors from a single
measurement.

A family of commuting Hermitian projectors on a finite-dimensional
space has a common orthogonal eigenspace decomposition. Let
$\{E_\lambda\}$ be its nonzero joint-eigenspace projectors on
$P\HH$. These subspaces are uniquely determined by the outcomes of
all exact shared readouts. There are at most $\dim(P\HH)$
nonzero subspaces. Consequently a finite list of shared projectors,
call it $E_1,\ldots,E_N$, suffices to distinguish all the joint
eigenspaces. Each $E_\lambda$ is the projector onto one definite
binary pattern of eigenvalues of this finite list.

It remains to show that this common refinement can itself be read
locally. For each $j$, select a local projector $\Pi_j^{(a)}$ in
each laboratory whose restriction to $P\HH$ is $E_j$, as in
Eq.~\eqref{eq:common-compression}. Define the Hermitian local
observable
\begin{equation}
B_a=\sum_{j=1}^N2^{j-1}P_j^{(a)},
\qquad \Pi_j^{(a)}=P_j^{(a)}\otimes\I_{\bar a}.
\label{eq:finest-readout}
\end{equation}
Each $B_a\otimes\I_{\bar a}$ commutes with $\rho$ because every
summand does. On $P\HH$ its eigenvalue is the binary code of the
joint eigenvalues of $E_1,\ldots,E_N$. Different occupied sectors
have different codes, and the code is identical for all observers.
Measuring the spectral projectors of $B_a$ therefore identifies the
same sector $\lambda$ in every laboratory without disturbing $\rho$.
Any eigenvalues occurring only outside $P\HH$ can be assigned to an
arbitrary existing outcome. This produces a complete projective
readout of the finest shared sectors.

Every projector of every other exact shared readout is a sum of
some $E_\lambda$, by their construction as common joint
eigenspaces. Thus all such records are coarse-grainings of this
unique finest one, up to relabeling. Since all $E_\lambda$ commute
with $\rho$, the sectors give Eq.~\eqref{eq:finest-sectors}.

\subsection{Proof of the full-rank corollary}

If $\rho$ is full rank, every nonzero positive projector $T$ has
$\Tr(\rho T)>0$. Suppose two different labels $x$ and $y$ both
have nonzero probability in an exact shared readout. Choose any
two distinct laboratories $a$ and $b$. Then $P_x^{(a)}$ and
$P_y^{(b)}$ are nonzero, so their product on the full Hilbert space
is a nonzero projector. Consequently
\begin{equation}
\Tr\!\left(\rho\Pi_x^{(a)}\Pi_y^{(b)}\right)>0,
\qquad x\ne y,
\end{equation}
contrary to the perfect-agreement requirement in
Eq.~\eqref{eq:operational}. This proves Corollary~\ref{cor:fullrank}.

\section{Robustness and branching-state estimates}\label{app:trace}

\subsection{Fixed-readout bounds and stability}

Let $\tau=\Delta(\rho)$ and write $p=\Tr(Q\tau)=1-\epsilon$. The equality of the probabilities follows because $Q$ is a sum of complete measurement blocks, so $\Delta$ does not change its expectation. Also $[Q,\tau]=0$. If $p>0$, define
\begin{equation}
\sigma_*=\frac{Q\tau Q}{p}\in\SSS_Q.
\label{eq:sigmastar}
\end{equation}
The difference $\tau-\sigma_*$ is block diagonal with respect to $Q\HH\oplus(\I-Q)\HH$. On the first block it is negative and has trace $-\epsilon$. On the second it is positive and has trace $\epsilon$. Thus $\|\tau-\sigma_*\|_1=2\epsilon$. The triangle inequality gives
\begin{equation}
d_Q(\rho)\leq\|\rho-\sigma_*\|_1
\leq\delta+2\epsilon.
\end{equation}
This explains why the upper error is linear in $\epsilon$: the pinched state has no off-diagonal terms connecting the matching and mismatching outcomes. If we instead projected an arbitrary coherent state directly, a standard gentle-measurement estimate would yield a weaker square-root correction.

For the lower bound, take any $\sigma\in\SSS_Q$. By definition $\Delta(\sigma)=\sigma$. Contractivity of the trace norm under the quantum channel $\Delta$ gives
\begin{equation}
\begin{aligned}
\delta&\leq\|\rho-\sigma\|_1
 +\|\Delta(\rho)-\Delta(\sigma)\|_1\\
&\leq2\|\rho-\sigma\|_1.
\end{aligned}
\end{equation}
Furthermore $\Tr[(\I-Q)\sigma]=0$. Using the variational characterization of trace distance for this binary test gives $2\epsilon\leq\|\rho-\sigma\|_1$. Taking the infimum over $\sigma$ proves Eq.~\eqref{eq:twosided}. The equivalence between $\delta=\epsilon=0$ and $\rho\in\SSS_Q$ follows directly from $\rho=\Delta(\rho)$ and support on $Q\HH$.

For the noisy-bit family, $\rho_\eta$ is diagonal in the local measurement basis, so $\delta=0$. The states $01$ and $10$ each occur with probability $\eta/4$, giving $\epsilon=\eta/2$. The normalized matching-state projection in Eq.~\eqref{eq:sigmastar} has trace-norm distance $\eta$ from $\rho_\eta$. The lower bound $d_Q\geq2\epsilon=\eta$ then proves $d_Q=\eta$. This also shows that the factor multiplying $\epsilon$ in the general upper bound cannot be made smaller. At $\eta=0$, the state has two finest shared labels. For every $\eta>0$, no nontrivial shared label exists.

To see perturbative stability, suppose $s=\|\rho-\rho'\|_1$ for two states tested with the same local projectors. Then
\begin{equation}
\begin{split}
|\delta(\rho)-\delta(\rho')|
&\leq\|(\rho-\rho')-\Delta(\rho-\rho')\|_1\leq2s,\\
|\epsilon(\rho)-\epsilon(\rho')|
&=|\Tr[Q(\rho-\rho')]|\leq s/2.
\end{split}
\label{eq:stability}
\end{equation}
The second inequality uses the fact that $Q$ is a projector and $\rho-\rho'$ has trace zero. The combined upper-bound expression $\delta+2\epsilon$ therefore changes by at most $3s$.

\subsection{Pointer dephasing and fragment-readout estimate}

Let $\tau=\Delta_S(\rho_{S\bm F})$ and let $Q$ be the projector onto identical pointer and fragment readouts. Because $\tau$ is block diagonal in the pointer basis and $Q$ includes that basis, the normalized projected state $\sigma=Q\tau Q/(1-r)$ has no coherence between distinct shared pointer values. It therefore has the sector structure required of an ideal shared record.

The gentle measurement inequality \cite{Winter1999} for the projector $Q$ gives $\|\tau-Q\tau Q\|_1\leq2\sqrt r$. Normalizing the successful outcome changes its trace norm by exactly $r$:
\begin{equation}
\left\|Q\tau Q-\frac{Q\tau Q}{1-r}\right\|_1=r.
\end{equation}
Together with the definition $D_{\mathrm{coh}}=\|\rho_{S\bm F}-\tau\|_1$, the triangle inequality yields Eq.~\eqref{eq:branch}.

For Eq.~\eqref{eq:unionbound}, condition on pointer value $x$. If all observed fragments return the correct $x$, then the joint matching projector succeeds. The event that this does not happen is the union of the individual failure events. Since the local projectors act on different fragments, their outcomes have an ordinary joint probability distribution, whether or not the fragment state factorizes. Summing the usual union bound over $x$, weighted by $p_x$, gives $r\leq\sum_ar_a$.

\section{Fixed-pointer comparisons and consensus bounds}\label{app:pointerproof}

\subsection{From the entropic SQD criterion to the pairwise form}

We spell out why the diagonal-pointer hypothesis in
Sec.~\ref{sec:pointerhierarchy} is needed. Let $X$ be a complete
rank-one eigenbasis of $\rho_S$ and consider one fragment $F_a$.
Dephasing $S$ in this basis gives
$\widetilde\rho_{SF_a}=\sum_xp_x|x\rangle\langle x|_S
\otimes\tau_x^{(a)}$. Its mutual information is precisely
$\chi(X:F_a)$. As dephasing leaves both $\rho_S$ and
$\rho_{F_a}$ unchanged under the eigenbasis assumption,
\begin{equation*}
I(S:F_a)-\chi(X:F_a)
 =H(\widetilde\rho_{SF_a})-H(\rho_{SF_a})\geq0.
\end{equation*}
The entropy increase under projective dephasing vanishes only when
$\rho_{SF_a}=\widetilde\rho_{SF_a}$. Moreover,
$\chi(X:F_a)=H(X)$ if and only if the occupied conditional fragment
states $\tau_x^{(a)}$ have mutually orthogonal supports: a classical
label can then be identified perfectly from that fragment. Since
$H(X)=H(S)$ here, the equalities in Eq.~\eqref{eq:sqdtest}
therefore imply Eq.~\eqref{eq:pairwisebroadcast}. Conversely, that
pairwise form makes the pointer perfectly readable and saturates all
three equalities. Without a diagonal pointer marginal, the entropic
equalities alone need not specify a complete readout of $X$.

\subsection{Pointer-overlap identity and conditional independence}

Fix the complete rank-one pointer basis $X=\{|x\rangle_S\}$ of Sec.~\ref{sec:pointerhierarchy}, with $\rho_S$ diagonal in $X$, and retain only labels with $p_x=\Tr(\rho\,|x\rangle\langle x|_S)>0$. Suppose first that $\rho$ has the shared pointer record $\mathrm R_X$. By nondisturbance of the local system pointer measurement, the complete state is block diagonal:
\begin{equation}
\rho=\sum_xp_x|x\rangle\langle x|_S\otimes\tau_x.
\label{eq:proofcq}
\end{equation}
The fragment readout projectors $P_x^{(a)}$ identify $x$ with certainty, so the support of $\tau_x^{(a)}=\Tr_{\bar a}\tau_x$ lies in the range $\HH_{F_a,x}$ of $P_x^{(a)}$. The projectors have orthogonal ranges for different $x$. Thus every $\rho_{SF_a}$ has the ideal form in Eq.~\eqref{eq:pairwisebroadcast}, and hence is SQD with the common pointer $X$. This proves $\mathrm R_X\subseteq\mathrm{SQD}_X\cap\mathrm{ND}_X$.

Conversely, suppose $\rho\in\mathrm{SQD}_X\cap\mathrm{ND}_X$. Nondisturbance of the system pointer again gives Eq.~\eqref{eq:proofcq}. For each $a$, SQD gives mutually orthogonal supports $\HH_{F_a,x}=\supp\tau_x^{(a)}$. The positivity of $\tau_x$ implies that it has no component outside the range of $P_x^{(a)}$ on fragment $a$: indeed
\begin{equation}
\Tr[(\I-P_x^{(a)})\tau_x^{(a)}]=0
\quad\Longrightarrow\quad
(\I-P_x^{(a)})\tau_x=0.
\end{equation}
(The implication follows because a positive operator has zero weight in a subspace only when its rows and columns vanish there.) As this holds for all $a$, $\supp\tau_x\subseteq\bigotimes_a\HH_{F_a,x}$. Reading the indicated local support projectors then perfectly agrees with the system pointer and commutes with the entire state. Therefore $\rho\in\mathrm R_X$, proving Eq.~\eqref{eq:overlapidentity} and the global form \eqref{eq:generalpointer}.

Within this class, conditional fragment independence is equivalent to $\tau_x=\bigotimes_a\tau_x^{(a)}$ for all $p_x>0$. The identity
\begin{equation}
D\!\left(\tau_x\Big\|\bigotimes_a\tau_x^{(a)}\right)
 =\sum_a H(\tau_x^{(a)})-H(\tau_x)
\end{equation}
follows by expanding the logarithm of a tensor product. Positivity of relative entropy, with equality only for identical states, then establishes Eq.~\eqref{eq:conditionaltotal} and the equivalence of $J=0$ with SBS within $\mathrm R_X$. In particular, the proof makes no assumption of pairwise or higher-order independence before setting $J=0$.

\subsection{Derivation of the geometric and accessible-information bounds}

Let the fixed readout include the complete rank-one system pointer $X$, and let $Q_x=|x\rangle\langle x|_S\otimes\bigotimes_aP_x^{(a)}$. Every state $\sigma\in\SSS_Q$ satisfies $\sigma=\sum_x Q_x\sigma Q_x$ and is therefore of the form in Eq.~\eqref{eq:generalpointer}. In particular, $\SSS_Q\subseteq\mathsf{SQD}_X$, where the SQD family allows the choice of orthogonal fragment support spaces to vary. Taking the infimum over the larger family gives $\inf_{\omega\in\mathsf{SQD}_X}\|\rho-\omega\|_1\leq d_Q(\rho)$. The upper bound in Eq.~\eqref{eq:sqdgeometric} is Theorem~\ref{thm:robust}.

For any local fragment measurement whose calibrated result $Y_a$ has error rate $e_a=\Pr(X\ne Y_a)$, introduce the binary indicator of the error event. Knowing $Y_a$ and this indicator leaves at most $\log_2(k-1)$ further bits to identify an incorrectly reported value of $X$. Therefore the elementary Fano argument yields
\begin{equation}
H(X|Y_a)\leq h_2(e_a)+e_a\log_2(k-1).
\end{equation}
Subtracting from $H(X)$ gives the second inequality in Eq.~\eqref{eq:fanobridge}. The first holds because the accessible mutual information maximizes over local measurements. Measuring $X$ on $S$ and the displayed $Y_a$ on its fragment is one allowed choice. Moreover, the event $X\ne Y_a$ is contained in the event that not all observers agree, so $e_a\leq\epsilon$.

If $\rho_S$ is diagonal in the rank-one pointer basis, $H(X)=H(S)$. The function $f_k(u)=h_2(u)+u\log_2(k-1)$ is increasing on $0\leq u\leq1-1/k$. If $\delta_\pi+2\epsilon_\pi\leq t$ for any set of candidate local readouts on a partition $\pi$, then $e_a\leq\epsilon_\pi\leq t/2$, and therefore
\begin{equation}
I_{\rm acc}(S:F_a^\pi)\geq H(S)-f_k(t/2)
\quad\text{for every }a.
\end{equation}
Thus, for each partition separately, shared-record success implies accessible-information success with the indicated threshold. Taking probabilities over any $\mu_m$ gives Eq.~\eqref{eq:probabilitybridge}. Taking maximum partition sizes gives the corresponding redundancy implication. The reverse is false because controlling the measured information does not control the global coherence removed by unread measurements, as the GHZ example shows.

\section{Counterexamples for the fixed-pointer hierarchy}
\label{app:counterexamples}

The five examples below realize the regions marked A--E in
Fig.~\ref{fig:framework}. Unless otherwise stated, the tested
partition consists of the designated system $S$ and two disjoint
fragments $F_1,F_2$.

\paragraph{A: A shared event need not encode all of $S$.}
Let $S=LQ$ contain a label qubit $L$ and an independent maximally
mixed qubit $Q$. With two label fragments, consider
\begin{equation}
\rho_A=\frac12\sum_{x=0}^1|x\rangle\langle x|_L
\otimes\frac{\I_Q}{2}\otimes
|xx\rangle\langle xx|_{F_1F_2}.
\label{eq:exampleA}
\end{equation}
The label is shared without disturbance, whereas the fragments
contain no information about $Q$:
\begin{equation}
H(S)=2,\qquad I(S:F_a)=I_{\rm acc}(S:F_a)=1.
\label{eq:exampleAnumbers}
\end{equation}
Consequently $\rho_A\in\mathrm R$, but QD and SQD fail for the
complete pointer of $S$.

\paragraph{B: QD without a shared fact.}
Let $S=S_1S_2$, let
$\Phi=|\Phi^+\rangle\langle\Phi^+|$ denote a Bell projector,
and take
\begin{equation}
\rho_B=\Phi_{S_1F_1}\otimes\Phi_{S_2F_2}.
\label{eq:exampleB}
\end{equation}
Each fragment satisfies $I(S:F_a)=H(S)=2$, but
$I_{\rm acc}(S:F_a)=1$. Thus QD holds whereas SQD fails. The
complete state is pure and admits no nontrivial exact shared event.

\paragraph{C: SQD without global nondisturbance.}
For $S,F_1,F_2$, take the coherent GHZ state
\begin{equation}
|G\rangle=(|000\rangle+|111\rangle)_{SF_1F_2}/\sqrt2,
\qquad\rho_C=|G\rangle\langle G|.
\label{eq:exampleC}
\end{equation}
Each $S:F_a$ marginal is a classical copy of the system $Z$ label:
\begin{equation}
I(S:F_a)=I_{\rm acc}(S:F_a)
=\chi(X:F_a)=H(S)=1.
\label{eq:exampleCnumbers}
\end{equation}
Yet an unread pointer measurement removes the
$|000\rangle\langle111|$ coherence. Hence
$\rho_C\in\mathrm{SQD}_X\setminus\mathrm R_X$, and the complete
pure state has no nontrivial exact shared event. Tracing out $S$
first produces a different tested state with a shared classical bit,
illustrating the importance of the system boundary.

\paragraph{D: An exact pointer record need not be SBS.}
Let $F_a=L_aK_a$, where $L_a$ stores the pointer label and $K_a$
is a quantum register. Put
$|\phi_\theta\rangle=\cos\theta|00\rangle+\sin\theta|11\rangle$
on $K_1K_2$ for $0\leq\theta\leq\pi/4$, and define
\begin{equation}
\begin{split}
\rho_D(\theta)=\frac12\sum_{x=0}^1
 &|x\rangle\langle x|_S\otimes
 |xx\rangle\langle xx|_{L_1L_2}\\[-2pt]
 &\otimes|\phi_\theta\rangle\langle\phi_\theta|_{K_1K_2}.
\end{split}
\label{eq:exampleD}
\end{equation}
Each $L_a$ reads $X$ without disturbance, so
\begin{equation}
I(S:F_a)=I_{\rm acc}(S:F_a)=H(S)=1.
\label{eq:exampleDnumbers}
\end{equation}
For $\theta>0$, the conditional state on $K_1K_2$ is entangled,
and $\rho_D(\theta)\in\mathrm R_X\setminus\mathrm{SBS}_X$.
For $\theta=0$, it is SBS. The conditional-correlation measure
in Eq.~\eqref{eq:conditionaltotal} gives
\begin{equation}
J(F_1:F_2|S)=2h_2(\sin^2\theta),
\label{eq:exampleDtotal}
\end{equation}
which ranges from zero to two bits. Each conditional $K_1K_2$
state is pure, while its one-qubit reductions have eigenvalues
$\cos^2\theta$ and $\sin^2\theta$. Thus exact pointer sharing
can coexist with nonzero conditional entanglement.

\paragraph{E: A shared fact and QD need not imply SQD.}
Add a copied classical tag to the two Bell pairs of region B.
Let $S=LS_1S_2$ and $F_a=L_aK_a$, with
\begin{equation}
\begin{aligned}
\rho_E&=\rho_{\mathrm{tag}}^{LL_1L_2}\otimes
\Phi_{S_1K_1}\otimes\Phi_{S_2K_2},\\
\rho_{\mathrm{tag}}^{LL_1L_2}&=
\tfrac12\sum_{x=0}^1|xxx\rangle\langle xxx|.
\end{aligned}
\label{eq:exampleE}
\end{equation}
The tag $L$ is an exact shared event, while
$H(S)=I(S:F_a)=3$ and $I_{\rm acc}(S:F_a)=2$.
Hence $\rho_E\in\mathrm R\cap\mathrm{QD}$ but does not
satisfy SQD for the complete pointer.

% Place references_revised.bib alongside this .tex file to compile citations.
\bibliography{references}
\end{document}